\documentclass[twocolumn,prd,superscriptaddress]{revtex4}
\usepackage[pdftex]{graphicx}
\usepackage{amsmath,amsfonts,amssymb}
\usepackage{slashed}
\usepackage[T1]{fontenc}
\usepackage{xcolor}
\usepackage{dutchcal}
\newcommand{\be}{\begin{equation}}
\newcommand{\ee}{\end{equation}}
\newcommand{\B}[1]{\mathbf{#1}}
\newcommand{\BS}[1]{\boldsymbol{#1}}
\newcommand{\C}[1]{\mathcal{#1}}
\usepackage{multirow}
\usepackage{ulem}
\graphicspath{{figures/}}

\begin{document}
\title{Quantum interactions between a laser interferometer and gravitational waves}
\author{Belinda Pang}
\affiliation{Theoretical Astrophysics and Walter Burke Institute for Theoretical Physics, M/C 350-17, California Institute of Technology, Pasadena, California 91125}
\author{Yanbei Chen}
\affiliation{Theoretical Astrophysics and Walter Burke Institute for Theoretical Physics, M/C 350-17, California Institute of Technology, Pasadena, California 91125}

\begin{abstract}
LIGO's detection of gravitational waves marks a first step in measurable effects of general relativity on quantum matter. In its current operation, laser interferometer gravitational-wave detectors are already quantum limited at high frequencies, and planned upgrades aim to decrease the noise floor to the quantum level over a wider bandwidth. This raises the interesting idea of what a gravitational-wave detector, or an optomechanical system more generally, may reveal about gravity beyond detecting gravitational waves from highly energetic astrophysical events, such as its quantum versus classical nature. In this paper we develop a quantum treatment of gravitational waves and its interactions with the detector. We show that the treatment recovers known equations of motion in the classical limit for gravity, and we apply our formulation to study the system dynamics, with a particular focus on the implications of gravity quantization. Our framework can also be extended to study alternate theories of gravity and the ways in which their features manifest themselves in a quantum optomechanical system.
\end{abstract}

\maketitle

\section{Introduction}
With LIGO's detection of gravitational waves \cite{GW150914}, there's been interest in using gravitational wave detectors (including e.g. VIRGO \cite{virgo}, KaGRA \cite{kagra}) to study not only astrophysical sources, but the nature of gravity itself, including modified theories \cite{GW170814,nontensorialGWpulsar,yunes2016} and quantum gravity \cite{camelia1999,GW170104,arun2009}. There are also important questions related to the quantum nature of the LIGO probe both in terms of the implications for its sensitivity as a measurement device \cite{miao2017,downes2017}, as well as the possibility of it being a test bed to study the interplay of quantum mechanics and gravity.

However, up to this point, the quantum interaction between gravitational waves and a LIGO-like (optomechanical) system has not been carefully studied from a general relativistic point of view, despite the interest in using optomechanical systems to study low energy gravity effects. To date, the literature comprises mainly of theoretical studies of its interaction with classical Newtonian gravity \cite{kafri2014} and phenomenological models of quantum or semiclassical gravity \cite{kleckner2008,yang2013,pikovski2012, belenchia2016}. There are also general relativistic quantum formulations of weak gravity interactions with bosonic fields \cite{anastopoulos2013, blencowe2013, oniga2016}, but these treatments do not easily extend to more complex matter systems. In particular, Oniga et. al \cite{oniga2016bound} derived the master equation for a bound scalar field, similar to an optical cavity, but in crucial contrast to LIGO's operation did not allow for the boundary length to change. Furthermore, because these treatments focus on the decoherence to quantum matter, they take the view of the gravitational field as an equilibrated bath in which the effects of interaction perturbations cannot be observed. 

In this paper, we develop a canonical formulation of linear quantum gravity from Einstein's theory of general relativity interacting with a quantum LIGO-like system (probe) in processes involving gravitational waves (GWs), which in principle can be extended to study interactions of quantum LIGO with GWs from other gravitational theories derivable from the action principle, such as scalar-tensor theories \cite{damour1992}. In contrast to both the conventional role of GWs as a classical and predetermined signal in LIGO, as well as treatments of quantum gravity as a thermal bath coupled to quantum matter, our formulation treats both the matter probe and the GW field on equal footing as dynamical degrees of freedom in an enlarged Hilbert space. Importantly, this treatment allows the matter probe and GW field to act mutually on each other, as compared to the previous scenarios. The paper focuses their dynamics, and in particular examines the testable physical implications of GW quantization. We find that the the probe$\rightarrow$GW field direction of interaction recovers Einstein's field equations for the generation of GWs but where the stress-energy tensor $T_{\mu\nu}$ is quantum. Conversely, the GW$\rightarrow$probe direction of interaction recovers the same equations for LIGO's output field in the presence of a classical GW signal as was calculated previously \cite{kimble2001}. 

That the formalism in the classical gravity limit recovers well known equations of GW detection and generation provides a check on its validity, but it additionally predicts effects for which the quantum nature of the GW field becomes essential. Specifically, we find that the mutual interaction leads to quantum coherent backaction effects on the probe in such a way that requires the presence of quantum GW fluctuations in order to preserve commutation relations. Furthermore, these backaction effects can be shown to be analogous to the radiation reaction damping that comes from classical corrections to the Newtonian potential \cite{MTW}. This suggests that in order to include the effects of weak GR corrections to Newtonian potential on quantum matter consistently and without violating canonical quantization, those perturbations must themselves be quantum. 

As an interpretational tool, our formulation offers an alternative (though physically equivalent) perspective of the detection process -- conventionally, LIGO's GW detection is viewed in the Newtonian gauge, where the GW signal is understood to act as a strain force on LIGO's cavities' mirrors (test mass), whose motion then modulates the field inside the cavity. In our formulation in the TT gauge the GWs interact directly with the cavity field. While all measurable quantities are the same in either gauge, the latter facilitates an intuitive and straightforward derivation of the probe's ultimate quantum-limited measurement sensitivity, known as its quantum Cramer Rao bound (qCRB) \cite{helstromQuantumEstimation}, for which the test mass dynamics is shown to be irrelevant. This bound has interesting relations with GW radiation and decoherence, which will be discussed in an accompanying paper.

The paper is divided into the following sections: section \ref{description} provides a description of the physical system along with definitions and notations used in this paper; section \ref{framework} expounds our theoretical framework whereby we develop a Hamiltonian formulation of the interacting system, which we apply in sections \ref{GWDynamics} and \ref{probeDynamics} to separately study the GW field and probe dynamics in the presence of their mutual interaction.

\section{Description of System}\label{description}
In this section, we shall describe a second-generation laser interferometer gravitational-wave detector, like those in LIGO, VIRGO and KaGRA. 

\subsection{The Optomechanical System}

Let us consider a Michelson interferometer containing a Fabry-Perot cavity in each of its two arms, which additionally has a power recycling mirror to increase the power circulating inside the arm cavities, as well as a signal recycling mirror to increase the bandwidth of detection~\cite{GW150914}. 

It has been shown that the antisymmetric mode of the interferometer which carries the GW signal and the quantum noises, including the power and signal recycling enhancements, can be mapped to a single detuned Fabry-Perot cavity with an effective input mirror and a perfectly reflective end mirror~\cite{buonanno2003}. This introduces errors of $O(l_{\rm SRC}/L)$ for signal recycling cavity length $l_{\rm SRC}$ and cavity arm length $L$, and so the assumption is valid when $l_{\rm SRC}\ll L$ which is the case in experiment. 

Additionally, for simplicity we attribute the effects of radiation pressure to the end mirror alone, which is possible if we assume that the input mirror is infinitely massive, or by reducing the end mirror to half its actual mass which introduces errors of $\max\{\Omega L/c,T\}$ for input transmissivity $T$ \cite{buonanno2003}. In this way, we can assume that the input mirror falls along its geodesic. Then, in the traceless-transverse gauge, if we choose our coordinate frame so that its origin coincides with the position of the input mirror at some point along its worldline and its coordinate velocity is initially zero, then the coordinates of the input mirror are fixed in time.  

Finally, we choose the cavity axis to be the $x$-direction along which we constrain the mirror motion described by its center of mass coordinate, thereby allowing us to model the mirror as massive point particle (valid for $t_m\ll \lambda_{GW}$ for mirror thickness $t_m$). In reality the LIGO mirrors are suspended pendulums, but the error in making this assumption is $O(q/l_p)$ where $q$ is mirror displacement due to GW and radiation pressure and $l_p$ is the pendulum length. In summary, with the stated errors, the signal and quantum noise analysis for the LIGO Michelson interferometer can be mapped onto that for a single one-dimensional Fabry-Perot cavity and all the radiation pressure effect attributed to the end mirror. We choose the origin of our coordinate system to be the position of the input mirror, which means that it is also not affected by metric peturbations and we can therefore hold this coordinate fixed. We now have mapped LIGO to a basic optomechanical system~\cite{chen2013}.

\begin{figure}[h]\label{setup}
\includegraphics[scale=0.4]{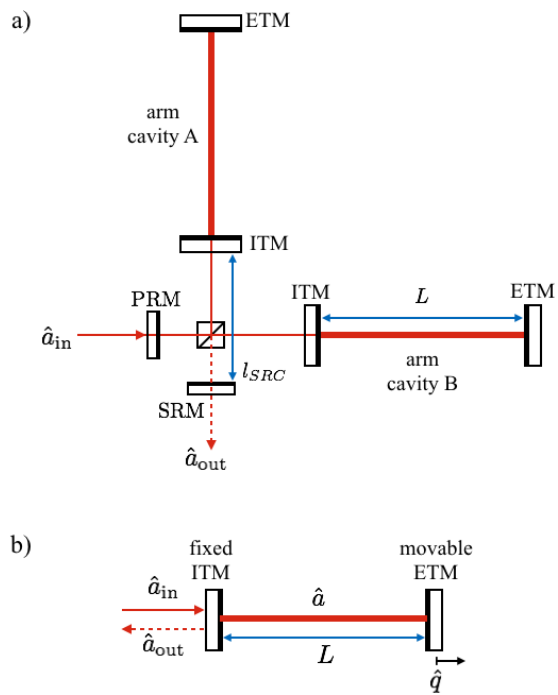}
\centering
\caption{Schematic of a second-generation laser interferometer gravitational-wave detector. Figure a) shows the full Michelson interferometer in its current configuration with power and signal recycling mirrors (PRM and SRM) and the two Fabry-Perot arm cavities. Here $L$ denotes the length of the arm cavity and $l_{SRC}$ denotes the length of the signal recycling cavity (shown here not to scale). The arm cavities' input mirror (ITM) has transmissivity T and its end mirror (ETM) is perfectly reflective with $R=1$. For low frequencies $\Omega$ of the GW wave such that $\Omega L/c\ll 1$ and for $T\ll1$, $l_{SRC}\ll L$ , the quantum inputs and outputs of the schematic in figure a) can be mapped to those of a single one-dimensional Fabry-Perot cavity shown in figure b).}
\end{figure}

\subsection{Inclusion of Gravity}

To describe space-time geometry, we assume weak metric perturbations about flat spacetime, such that the metric is given by $g_{\mu\nu}=\eta_{\mu\nu}+h_{\mu\nu}$, where $\eta_{\mu\nu}$ is the Minkowski metric with signature $(-+++)$ and $h_{\mu\nu}$ a small perturbation which we treat up to linear order. Although LIGO operates in earth's weak gravitational field, this is a constant longitudinal component whose effect on the test masses are balanced directly by the tension in the pendulums.  Since we only consider linear gravitational perturbation, it is safe to ignore earth's gravity because it does not couple to gravitational-wave contributions to $h_{\mu\nu}$ at this order.

\subsection{Terminology}

Throughout this paper and unless otherwise noted, Greek indices $\mu,\nu\,\rho$ etc. (with the exception of $\lambda$) denote spacetime components of vectors and tensors; English alphabet indices $i,j,k$ etc. denote purely spatial components; $\lambda$ denotes graviton polarizations; boldface form $\B{v}$ represents 3-vectors and $\vec{v}$ represents 4-vectors. We also use Einstein subscript summation notation where contraction is with respect to the background Minkowski metric (e.g $M_{\mu\nu}M^{\mu\nu}$). Repeated spatial indices denotes summation regardless of upper or lower position (e.g $v_iv_i$).

For terminology, ``optical mode'' refers to the optical field inside the Fabry-Perot cavity; ``test mass'' refers to the end mirror of the cavity; ``probe'' refers to the optomechanical system comprising of the optical mode, the test mass, and their interaction; ``system'' without a modifier refers to the GW field and the probe together; ``pump field'' refers to the optical input to the cavity and ``output field'' refers the cavity output on which measurement is performed.

\section{Theoretical Framework}\label{framework}
To study the interaction of weak gravity with macroscopic matter systems in finite time, we can use the  canonical quantization formulation, even though a full quantum theory of gravity is not yet available~\cite{kiefer2012}. Canonical quantization can be quite straightforward in systems with well-defined physical coordinates and velocities that appear in the Lagrangian in quadratic form, but here there are two difficulties. 

First is the fact that gravity has coordinate (or gauge) degrees of freedom which is mathematically reflected in the singularity of its Lagrangian. This means a physical state can be represented by multiple points that form a trajectory in phase space. Dirac is credited with developing the Hamiltonian formulation of such gauge theories, in which the degeneracies in phase space due to gauge degrees of freedom can be eliminated by to restricting to a hypersurface which itself is foliated by gauge orbits. Quantization can proceed in the usual way by ignoring the existence of gauge freedom, but physical quantum states must satisfy constraint conditions which ensure that the physical Hilbert space slices across the gauge orbits \cite{diracConstrainedHamiltonian}. Applying this approach to linearized gravity, Gupta derived a Hamiltonian for a pure gravitational field and the constraint conditions that must be satisfied by physical gravitational states. He also demonstrated that a pure gravitational field has only two physical gravitons, although more exist in virtual states in the presence of interaction \cite{guptaLinearGrav}. However, we can greatly simplify the quantization procedure by noting that, since we are only interested in studying leading order interactions involving incoming or outgoing gravitons, we can restrict the gravitational field to its two physical polarizations at the level of the action, eliminating its longitudinal and time-like components, which obviates the need for constraint conditions on physical states. While such a method would not give the correct result for interactions mediated by virtual gravitons, i.e. self-gravity of the detector, it is appropriate for studying leading order interactions between the interferometer and gravitational waves.

The second issue lies in obtaining the Hamiltonian for the optomechanical interaction between the test mass and the optical mode. To our best knowledge, to derive this interaction there currently only exist procedures which assume that the equations of motion for cavity are known {\it a priori}, whereupon a suitable Lagrangian producing those equations is constructed~\cite{lawCavity}. However, since our purpose is to study the unknown behaviour of an interacting system, the equations of motion must follow {\it from} the action instead of preceding it, and we develop an alternative approach so that the equations for all dynamical quantities of the system follow consistently from a canonical formulation beginning with the action.

\subsection{Gauge Fixing for Gravitational Field}
We begin with the linearized Einstein-Hilbert action for the metric in the harmonic gauge with $\partial_\mu h^{\mu\nu}=0$, and write 
\be\label{EHAction}
S_{EH}=-\frac{c^4}{32\pi G}\int d^4x \left[
\frac{1}{2}\partial_\mu h_{\alpha\beta}\partial^\mu h_{\alpha\beta}-\frac{1}{4}\partial_\mu h\partial^\mu h
\right]
\ee
where $h=h^\mu{}_{\mu}$ is the tensor trace. As discussed previously, to leading order the interactions between the interferometer and gravitational waves only involve the physical polarizations of the field, which we have the freedom to express in any gauge. Choosing the traceless-transverse (TT) gauge, we eliminate any time-like component of the field and expand its spatial components in the Fourier domain as
\be\label{hFourier}
h^{\rm TT}_{ij}(t,\B{x})=\int \frac{d^3\B{k}}{\sqrt{(2\pi)^3}}\tau_{ij}^\lambda(\B{k})h_\lambda(t,\B{k})e^{i\B{k}\cdot\B{x}}
\ee
where the index $\lambda=+,\times$ denotes the polarization. The tensor $\tau_{ij}^\lambda(\B{k})$ is the unit tensor for the $\B{k}$-mode component, and satisfies orthogonality, transverse, and traceless conditions:
\be
\tau^\lambda_{ij}\tau^{\lambda'}_{jk}=\delta_{\lambda,\lambda'}\delta_{ik},\quad
\B{k}\cdot\BS{\tau}^\lambda(\B{k})=0,\quad
{\rm Tr}[\BS{\tau}^\lambda]=0
\ee
Finally, the Einstein Hilbert action can be rewritten as
\begin{align}
S_{EH}&=\frac{c^4}{32\pi G}\int dt\int_{\frac{1}{2}} d^3\B{k}
\left[
\frac{1}{c^2}|\dot{h}_\lambda(t,\B{k})|^2-\B{k}^2|h_\lambda(t,\B{k})|^2
\right]\label{EHActionTT}\\
&\equiv \int dt\;\int_{\frac{1}{2}} d^3\B{k}\;\C{L}_{GW}^{(0)}(t,\B{k})\notag
\end{align}

We remark that $h$ is related to $h^*$ by
\be\label{complexConj}
h_\lambda^*(t,\B{k})\tau^\lambda_{ij}(\hat{\B{k}})=h_\lambda(t,-\B{k})\tau^\lambda_{ij}(-\hat{\B{k}})
\ee
Therefore, summing $h_\lambda\tau^\lambda_{ij}$ over all of $\B{k}$-space is physically equivalent to summing over $h_\lambda\tau^\lambda_{ij}$ and $h_\lambda^*\tau^\lambda_{ij}$ over half of k-space. The latter method allows us to treat $h_\lambda$ and $h_\lambda^*$ as independent degrees of freedom, in a similar approach to that of~\cite{tannoudjiPhotons} for the Hamiltonian formulation of electrodynamics.

\subsection{Optomechanical Interaction}
In this section, we briefly summarize the first principles derivation of the optomechanical interaction between the optical mode and test mass from the electromagnetic (EM) field action. The details of the derivations, including relativistic corrections and extension to multimodes, will be published in an accompanying paper (note that this derivation does not follow the work of Law \cite{lawCavity} because, due to the presence of other interactions, we cannot not posit \textit{a priori} equations of motion for the test mass and optical mode as is done in Ref.\cite{lawCavity}). For simplicity, we present the derivation in Minkowski space, although adding the metric perturbation to our analysis is straightforward. As we will show, the optomechanical interaction is hidden in the spatial boundary conditions of the EM field. With the appropriate coordinate transformation, the boundary condition appears as an explicit term in the action instead of being embedded in the integration limits. We consider the ideal case where the cavity has perfectly reflective boundaries $R=1$, although we will relax this assumption later on to allow transmission of the pump and output fields. 

Let us first write down the EM action in Minkowski space, denoting the EM vector potential by $\C{A}^{\mu}$. Since there are no charged currents, we can apply the Coulomb gauge and set the time component of the vector potential to zero, or $\C{A}^0=0$, and write
\be\label{EMActionMink}
S_{EM}^\eta=\frac{1}{2\mu_0}\int_{V_{\rm cav}} d^4x\left[
\frac{1}{c^2}\dot{\C{A}}_j^2-\left(\partial_i\C{A}_j\right)^2
\right]=\int dt\;L_{EM}^\eta
\ee
with the $\eta$ superscript denoting Minkowski space. Importantly, $V_{\rm cav}$ specifies the spatial limits of the EM field contained inside the Fabry-Perot cavity. We can approximate this field as having a constant mode profile transverse to the cavity axis, so that only the part of the field which propagates along the cavity axis is dynamical. The vector potential is then separable as $\C{A}_j(t,\B{x})=u(y,z)A_j(t,x)$, where $j$ can only take values of $y,z$ to satisfy the Coulomb gauge condition. Then, defining the mode volume $\C{U}=\int dy\int dz |u(y,z)|^2$, the Lagrangian $L_{EM}^\eta$ in Eq.~\eqref{EMActionMink} can be written as an integral along the cavity axis
\be\label{EMLagrangianBC}
L_{EM}^\eta=\frac{\C{U}}{2\mu_0}\int_0^{L'} dx\; \left[
\frac{1}{c^2}\dot{A}_j^2-\partial_x A_j^2\right]
\ee
where $L'$ is the coordinate length of the cavity, or equivalently the position of the test mass, and is a dynamical quantity. The optomechanical interaction derives from this dynamical boundary condition, which comes from the physical constraint that the EM field must vanish at the cavity's perfectly reflecting mirrors, such that $A_j(t,0)=A_j(t,L')=0$. We perform the following coordinate transformation so that the test mass position appears explicitly in the Lagrange density (note by coordinate transformation, we are referring to the coordinates which themselves and whose velocities appear in the Lagrangian, and not spacetime coordinates)
\be\label{coordTrans}
A_j(t,x)=\frac{c^2}{\omega_0 L'}\Phi_j(t)\sin(\kappa x),\quad \kappa=\frac{n\pi}{L'},\,n\in \mathbb{Z}
\ee
where $\omega_0=n\pi c/L$ is the resonant frequency of the cavity at its equilibrium length $L$. The coordinate transformation of Eq.~\eqref{coordTrans} automatically satisfies the boundary conditions, and separates the spatial part of the field, which is subject to a time dependent boundary condition, from the naturally time varying part. The coordinate $\Phi_j(t)$ is no longer a field defined at each point in spacetime, but represents excitations of the spatially extended optical mode. The collection of all modes $\Phi_j(t)$ with different values of $n$ contain the same information as $A_j(t,x)$, but in this paper we only consider a single mode, which is closest to the pumping frequency.

Then, defining $q=L'-L$ to be the motion of the test mass about equilibrium and substituting Eq.~\eqref{coordTrans} into Eq.~\eqref{EMLagrangianBC}, to leading order in $q/L$, we find $L_{EM}^\eta=L_{EM}^{(0)}+L_{OM}$, where
\begin{subequations}
\begin{align}
L_{EM}^{(0)}&=\frac{\C{U}}{2\mu_0}\left[
\frac{1}{2\omega_0}\dot{\Phi}_j^2(t)-\frac{\omega_0}{2}\Phi_j^2(t)
\right]\label{EMLagrangian}\\
L_{OM}&=\frac{\C{U}}{2\mu_0}\frac{\omega_0q}{L}\Phi_j^2(t)\label{OMLagrangian}
\end{align}
\end{subequations}
where summation over $j$ (through $y$ and $z$) is implied. The test mass position $q$ now appears explicitly in the Lagrangian, and we also find the optomechanical interaction term $L_{OM}$.

\subsection{Interaction between GW and Probe}
The probe consists of the optical mode and the test mass, whose actions in perturbed spacetime can be written as $S_{\rm probe}=S_{EM}^\eta+S_{EM}^h+S_q$, where $S_{EM}^\eta$ was the EM action in Minkowski space and was defined in Eq.~\eqref{EMActionMink} and $S_{EM}^h$ is the first order term in the expansion of $S_{EM}\propto -\int d^4x \sqrt{-g}\;g_{\alpha\mu}g_{\beta\nu}F^{\mu\nu}F^{\alpha\beta}$ with respect to $h_{\mu\nu}$, which we write below

\begin{align}
S_{EM}^{h}=&-\frac{1}{2\mu_0}\int_{V_{\rm cav}} d^4x\;h_{ij}\Bigg[
\frac{1}{c^2}\dot{\C{A}}_i\dot{\C{A}}_j-(\partial_i \C{A}_k)(\partial_j \C{A}_k)\notag\\
&-(\partial_k \C{A}_i)(\partial_k \C{A}_j)
\Bigg]=\int dt\;L_{EM}^{h}\label{EMActionH}
\end{align}

Similarly, we expand the action for the test mass $S_q=-mc\int d\tau$ to leading order in $h_{\mu\nu}$ and obtain
\be\label{testMassAction}
S_q=\frac{m}{2}\int dt\;\left[\dot{\B{x}}_q^2+h_{ij}(t,\B{x}_q)\dot{x}_q^i\dot{x}_q^j\right]
=\frac{m}{2}\int dt\;\dot{q}^2
\ee
where to obtain the second equality we've ignored terms of $O(v^2/c^2)$ as well as the test mass's degrees of freedom of motion in the $y,z$ directions, which are non-interacting and trivial. We write the corresponding Lagrangian as $L_q^{(0)}=m\dot{q}^2/2$.

Thus, we find that the interaction between the GW and the probe only concerns the EM field, and, denoting this interaction by $L_{GW}^{\rm int}$, we have $L_{GW}^{\rm int}=L_{EM}^{h}$. Substituting $h_{ij}$ in Eq.~\eqref{EMActionH} by its expansion into transverse-traceless Fourier modes given in Eq.~\eqref{hFourier}, and performing the same coordinate transformation to the EM vector potential as given in Eq.~\eqref{coordTrans}, we obtain

\begin{align}
L_{GW}^{\rm int}=&-\frac{\C{U}}{2\mu_0}\int_{\frac{1}{2}}d^3\B{k}
\left[
J^\lambda_{ij}(\B{k})h_\lambda(t,\B{k})+J^{\lambda *}_{ij}(\B{k})h_{\lambda}^*(t,\B{k})
\right]\notag\\
&\times\left[
\frac{1}{2\omega_0}\dot{\Phi}_i\dot{\Phi}_j-\frac{\omega_0}{2}\left(\Phi_i\Phi_j+\delta_{ix}\delta_{jx}\Phi_k^2\right)
\right]\label{LIntGW}
\end{align}
where $J^\lambda_{ij}(\B{k})$ is a GW mode profile function, and is given by
\be\label{GWModeProfile}
J^\lambda_{ij}(\B{k})=\frac{-i(e^{ik_x L}-1)}{k_xL}\frac{\tau^\lambda_{ij}(\B{k})}{{\sqrt{(2\pi)^3}}}
\ee

The factor of $-i(e^{ik_x L}-1)/k_xL$ derives from the variation of the gravitational wave over the spatial extent of the optical mode. For long wavelength GWs such as those from LIGO's astrophysical sources, $k_x L\ll 1$ and $J_\lambda(\B{k})$ simply reduces to the interacting polarization tensor component. However, in order to study backaction due to GW radiation from LIGO itself, one must include this factor to ensure convergence, since the backaction effect requires that radiated GW vary over the length of the cavity.

\subsection{Canonical Quantization}
The full Lagrangian for the system is then
\be\label{LFull}
L=L_q^{(0)}+L_{EM}^{(0)}+\int dt\int_{\frac{1}{2}}d^3\B{k}\;\C{L}_{GW}^{(0)}(t,\B{k})+L_{OM}+L_{GW}^{int}
\ee

We then proceed with canonical quantization, first performing a Legendre Transform to identify conjugate pairs $\{h_\lambda(\B{k}),\Pi^*_\lambda(\B{k})\},\,\{\Phi_i,\mathbb{P}_i\},\,\{q,p\}$. We remark that although the photon polarizations appear coupled in  $L_{GW}^{\rm int}$ in Eq.~\eqref{LIntGW}, the interaction between the different polarizations occur on much shorter timescales than those of interest. To show this, and also to express the Hamiltonian in more familiar variables, we perform the following canonical transformation on the EM conjugate pair by defining $\alpha^{(i)}_1,\alpha^{(i)}_2$ such that
\begin{subequations}
\begin{align}\label{canonicalTrans}
\Phi_i&=\sqrt{\frac{2\mu_0}{\C{U}}}\left[
\alpha^{(i)}_{1}\cos\omega_0 t+\alpha^{(i)}_{2}\sin\omega_0 t
\right]\\
\mathbb{P}_i&=-\sqrt{\frac{\C{U}}{2\mu_0}}\left[
\alpha^{(i)}_{1}\sin\omega_0 t-\alpha^{(i)}_{2}\cos\omega_0 t
\right]
\end{align}
\end{subequations}
where the superscript $(i)$ represents photon polarization. Anticipating that the optical mode is driven by a resonant pump field, here we have separated the fast time dependence of the pump which is rotating at $\omega_0$ in the tuned configuration, thereby going into the interaction picture with respect to the optical carrier frequency. In this co-rotating frame we find terms oscillating at $2\omega_0$. Since these effects occur on much shorter timescales than the interaction dynamics, for weak interactions we ignore them under the rotating wave approximation (RWA) \cite{wallsMilburn} and we find that the two photon polarizations interact with the test mass and GWs independently and in identical ways. We therefore suppress the photon polarization superscript, and obtain the Hamiltonian $H=H_q^{(0)}+H_{GW}^{(0)}+H_{OM}+H_{GW}^{\rm int}$. Here $H_q^{(0)}$ and $H_{GW}^{(0)}$ are the free Hamiltonians for the test mass and metric perturbation respectively, $H_{OM}=-\omega_0(\hat{\alpha}_1^2+\hat{\alpha}_2^2)\hat{q}/2L$ is the optomechanical interaction, and $H_{GW}^{\rm int}$ is the interaction between the GW field and the probe, given by
\be\label{HGWInt}
H_{GW}^{\rm int}=-\frac{\omega_0}{4}\left(\hat{\alpha}_{1}^2+\hat{\alpha}_{2}^2\right)
\int d^3\B{k} \; J_\lambda(\B{k})\hat{h}_\lambda(t,\B{k})
\ee
where we've used $J_\lambda$ to represent $J^\lambda_{xx}$. We point out that only the $xx$ component of the GW field interacts with our probe, where $x$ is the direction of propagation of the dynamical photons in the optical mode.

We then quantize canonically by imposing the commutation relations 
\begin{subequations}
\be\label{commutator}
[\hat{q},\hat{p}]=i\hbar,\quad
\left[\hat{\alpha}_1,\hat{\alpha}_2\right]=i\hbar
\ee
\be\label{commutator2}
\left[\hat{h}_\lambda(\B{k}),\hat{\Pi}^\dagger_{\lambda'}(\B{k}')\right]=\left[\hat{h}^\dagger_\lambda(\B{k}),\hat{\Pi}_{\lambda'}(\B{k}')\right]=i\hbar\delta_{\lambda\lambda'}\delta^2(\B{k}-\B{k}')
\ee
\end{subequations}
 
In order to perform measurement on the probe state while maintaining a constant amplitude inside the cavity, we must relax the perfect reflectivity condition to couple the probe to an external pump field whose ingoing photons drive the optical mode and whose outgoing photons are measured by the photodetector. The outgoing photons may be also thought of as ancillae which ensures that the probe state evolves unitarily during continuous measurement without measurement based feedback. We enlarge our Hilbert space to include the external pump by adding $H_{\rm ext}=i\sqrt{2\gamma}\left[\hat{a}^\dagger\hat{c}_{x=0}-\hat{a}\hat{c}^\dagger_{x=0}\right]
-i\int_{-\infty}^\infty dx\;\hat{c}^\dagger_x\partial_x \hat{c}_x$ to account for the interaction between the pump and probe (going directly to the interaction picture with respect to the free evolution of the pump field). Here $\hat{a}$, $\hat{a}^\dagger$ are raising and lowering operators of the optical mode, defined such that $\hat{\alpha}_1=\sqrt{\hbar/2}\left(\hat{a}+\hat{a}^\dagger\right)$ and $\hat{\alpha}_2=-i\sqrt{\hbar/2}\left(\hat{a}-\hat{a}^\dagger\right)$. We shall refer to $\hat \alpha_{1,2}$ as the amplitude and phase quadratures.

Relaxing the condition of perfect reflectivity also allows LIGO to operate in the detuned configuration where the pump field has off resonant frequency $\omega_L=\omega_0+\Delta$. The fast evolution in the canonical transformation in Eq.~\eqref{canonicalTrans} is then given by $\omega_L$ instead of the $\omega_0$, and this results in an additional term in the Hamiltonian given by $H_\Delta=-\Delta\left(\hat{\alpha}_1^2+\hat{\alpha}_2^2\right)/2$.

Assuming large average amplitude inside the cavity, we linearize the Hamiltonian by writing $\hat{\alpha}_1\rightarrow \bar{\alpha}+\delta\hat{\alpha}_1$ and $\hat{\alpha}_2\rightarrow \delta\hat{\alpha}_2$. Then keeping only terms linear in small quantity $\delta$ in the interaction terms (which are already small), we write down the final form of our Hamiltonian:

\begin{align}
H=&H_q^{(0)}+H_{GW}^{(0)}+H_{\rm ext}-\frac{\Delta}{2}\left(\hat{\alpha}_1^2+\hat{\alpha}_2^2\right)\notag\\
&-\omega_0\bar{\alpha}\hat{\alpha}_1
\left[
\frac{\hat{q}}{L}+\frac{1}{2}\int d^3\B{k} \; J_\lambda(\B{k})\hat{h}_\lambda(t,\B{k})
\right]\label{HFull}
\end{align}
where we take $\Delta\rightarrow 0$ to recover the tuned configuration. 

For notational simplicity we denote the integral over GW $\B{k}$-modes by $\hat{\C{h}}(t)$, or
\be\label{hKInt}
\hat{\C{h}}(t)\equiv\int d^3\B{k} \; J_\lambda(\B{k})\hat{h}_\lambda(t,\B{k})
\ee
 and point out that in the long wavelength approximation where $k_xL\ll 1$ we have $\hat{\C{h}}(t)=\hat{h}^{\rm TT}_{xx}(t,\B{x}=\B{0})$ according to Eq.~\eqref{hFourier}. 
  
For a strong excitation of the GW field from an astrophysical event such as a BBH merger, we can separate the GW field into a large classical component along with quantum fluctuations and write $\hat{\C{h}}(t)\rightarrow {h}_s(t)+\hat{\C{h}}(t)$, and the interaction Hamiltonian becomes
\be\label{HInt}
H_{\rm int}=-\frac{\omega_0\bar{\alpha}\hat{\alpha}_1}{2}\left[\C{h}_s(t)+\hat{\C{h}}(t)\right]
\ee

We also point out that $H_{\rm ext}$ was not derived from a fundamental action and was added phenomenologically in accordance with the standard formulation for input-output theory in quantum optics \cite{wallsMilburn}. The only concern here would be the interaction between the external pump and the GW field, but that interaction is negligible since the power in the pump without the amplification effects of a Fabry Perot cavity is orders of magnitude smaller than that of the optical mode. 

\subsection{Equations of Motion}\label{EOM}
We identify the input and output pump field from the external field operators $\hat{c}_x$ \cite{chen2013}
\be\label{inputOutputOperators}
\hat{a}_{\rm in}=\hat{c}_{x=0^-},\quad
\hat{a}_{\rm out}=\hat{c}_{x=0^+},\quad
\hat{c}_{x=0}=\frac{\hat{c}_{x=0^-}+\hat{c}_{x=0^+}}{2} 
\ee
and denote their corresponding amplitude and phase quadratures by $\hat{\alpha}_{\rm in}^{(1,2)}$. From the Hamiltonian in Eq.~\eqref{HFull} we derive the following Heisenberg EOM for the probe and the GW field
\be\label{testMassEOM}
\dot{\hat{q}}=p/m,\quad \dot{\hat{p}}=\omega_0\bar{\alpha}\hat{\alpha}_1/L
\ee
\begin{subequations}\label{cavEOM}
	\be\label{ampEOM}
		\dot{\hat{\alpha}}_1=-\Delta\hat{\alpha}_2-\gamma \hat{\alpha}_1+\sqrt{2\gamma}	\hat{\alpha}^{\rm in}_{1}
	\ee
	\begin{align}
		\dot{\hat{\alpha}}_2=&\,\Delta\hat{\alpha}_1-\gamma\hat{\alpha}_2+\sqrt{2\gamma}\hat{\alpha}^{\rm in}_{2}\notag\\
		&-\frac{\omega_0 \bar{\alpha}}{2}\left[\C{h}_s(t)+\int d^3\B{k}\;J_\lambda(\B{k})\hat{h}_\lambda(\B{k})\right]-\frac{\omega_0\bar{\alpha}}{L}\hat{q}\label{phaseEOM}
	\end{align}
\end{subequations}
\begin{subequations}\label{gravEOM}
	\begin{gather}
		\dot{\hat{h}}_\lambda(\B{k})=\frac{1}{M_G}\hat{\Pi}_\lambda(\B{k})\label{hEOM}\\
		\dot{\hat{\Pi}}_\lambda(\B{k})=-\omega_k^2M_G\hat{h}_\lambda(\B{k})+\frac{\omega_0\bar{\alpha}}{2}J_\lambda^*(\B{k})\hat{\alpha}_1\label{PiEOM}
	\end{gather}
\end{subequations}
where $\omega_k=c|\B{k}|$ and we've defined $M_G=c^2/32\pi G$.

\subsection{Discussion of Gauge Choice}
We point out that, as expected, in the TT gauge the metric perturbations do not affect the coordinate motion of a particle moving along a geodesic~\cite{MTW} to order $(v^2/c^2)$, as evident in Eq.~\eqref{testMassAction}. Additionally, if the particles are initially at rest (such that $v=0$), and experience no other forces, then linear metric perturbations will not affect coordinate motion for even appreciable values $v/c$. This is because the gauge symmetry for gravitational fields is a diffeomorphism, or a local symmetry, and so the coordinate length between two free falling particles is gauge dependent. However all physical quantities must gauge invariant, and in particular the time elapsed between when a photon enters the cavity to when it is reflected back (i.e. with respect to an observer sitting on the input mirror) is the same in either gauge, despite the difference between the two gauges in terms of the coordinate distance that the photon travels. It should be noted that for our system, the test mass in not actually a free falling particle since it interacts optomechanically with the optical mode. This means that its position coordinate may be affected by metric perturbations even in the TT gauge, and in fact are through the coupling of $L_{OM}$ to $h_{ij}$. However, this term is of $O(hq/L)$, which is second order in small quantities and therefore ignored in Eq.~\eqref{LIntGW}
\begin{figure}[h]\label{gauges}
\includegraphics[scale=0.45]{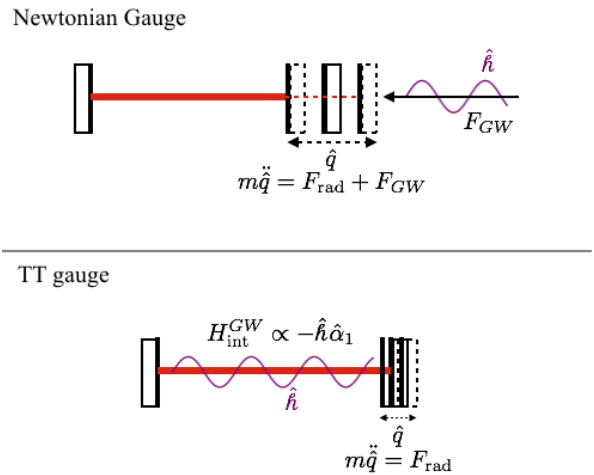}
\centering
\caption{Representations of the GW interaction in the Newtonian versus TT gauges. In the Newtonian gauge, the gravitational wave exerts a strain force $F_{GW}$ so the the test mass position is driven by both radiation pressure and gravitational wave forces. In contrast, in the TT gauge the GW interacts directly with the optical cavity mode and the test mass position is driven by the radiation pressure force alone. The two pictures are physically equivalent descriptions of the dynamics of a cavity whose mirrors fluctuate about their geodesic due to radiation pressure. In the presence of an incoming GW, geodesics of the two mirrors deviate and time delay for a photon entering to be reflected back changes. In the Newtonian gauge, the change in time delay is reflected in the test mass coordinate, while in the TT gauge this effect is directly accounted for by a phase shift in the cavity mode. The TT gauge viewpoint allows for a canonical description of the interaction.}
\end{figure}

The consequence of the TT gauge choice is that in this picture, the GWs interact directly with the optical mode, in contrast to the often-taken point of view that GWs exert a strain force onto the test mass whose motion then causes a phase shift \cite{kimble2001}. The TT gauge view allows for straightforward derivation of the qCRB for LIGO, or the fundamental limit to measurement sensitivity. We will discuss the details of this and its implications in an accompanying paper.

\section{Generation of Quantum Gravitational Waves}\label{GWDynamics}
Since our formalism treats the GW field and the probe on equal footing, the interaction between them is bidirectional, meaning that in addition to the Hamiltonian in Eq.~\eqref{HFull} describing how the probe evolves under GW interaction, it also governs how the probe affects the dynamics of the GW field. Specifically we recover the quantum analogue of the classical quadrupole moment formula as derived from Einstein's field equations \cite{MTW}:
\be\label{quadrupoleFormula}
\hat{h}_{ij}^{\rm TT}(t,\B{x})=\frac{2G}{c^4}\frac{\ddot{\hat{I}}_{ij}^{\rm TT}(t-|\B{x}|/c)}{|\B{x}|}
\ee
where $\hat{\B{I}}^{\rm TT}$ is the TT projection of the mass quadrupole moment. The result is unsurprising, but it serves as a demonstration of the equivalence in the relations between perturbative quantum gravity interacting with quantum matter and that of the classical scenario, specifically that quantum stress energy radiates GW in the same way as classical matter, and moreover that the radiation is quantum. This is in contrast to semiclassical gravity, which postulates that spacetime is classical while matter fields are quantum, such that the expectation value must be taken over $\ddot{\hat{\B{I}}}$. We emphasize that the semi-classical and fully quantum viewpoints are in nature different and indeed has testable physical implications, at least in principle. First, Eq.~\eqref{quadrupoleFormula} implies that non-classical states of quantum matter will result in non-classical states of GWs and second, the quantum generation of GWs will result in quantum coherent backaction effects onto matter, which has not been previously considered and will be discussed further in section \ref{probeDynamics}.

To obtain Eq.~\eqref{quadrupoleFormula} we solve the the set of coupled differential equations in Eqs.~\eqref{hEOM} and~\eqref{PiEOM} in the frequency domain. Defining the Fourier pair using the conventions
\be\label{FourierPair}
\hat{O}(\Omega)=\int_{-\infty}^{\infty} dt\; \hat{O}(t)e^{-i\Omega t},\quad
\hat{O}(t)=\int_{-\infty}^{\infty}\frac{d\Omega}{2\pi}\;\hat{O}e^{-i\Omega t}
\ee
We have
\be\label{hEOMFreq}
\tau_{ij}(\B{k})\hat{h}_\lambda(\Omega,\B{k})=\frac{\omega_0\bar{\alpha}}{2M_G}
\frac{\tau_{ij}(\B{k})J_\lambda(\B{k})}{(\omega_k^2-\Omega^2)}\hat{\alpha}_1(\Omega)
\ee
Since the probe is a Newtonian source, we can apply the slow motion condition and make the simplifying approximation that $|\B{k}|L\ll1$, so that $J_\lambda({\B{k}})\rightarrow\tau_{xx}(\B{k})/\sqrt{(2\pi)^3}$. Remembering that we are already in the TT gauge, the inverse spatial Fourier transform of Eq.~\eqref{hEOMFreq} will give us the gauge fixed GW field in configuration space, or $\hat{h}^{\rm TT}_{ij}(\Omega,\B{x})$. Then from our equations of motion we obtain
\be\label{GWField}
\hat{h}_{ij}^{\rm TT}(\Omega,\B{x})=\frac{1}{2M_G}\int\frac{d^3\B{k}}{(2\pi)^3}
\frac{\omega_0\bar{\alpha}\hat{\alpha}_1(\Omega)
\tau_{ij}^\lambda(\B{k})\tau_{xx}^\lambda(\B{k})e^{i\B{k}\cdot\B{x}}}
{\omega^2-\Omega^2}
\ee
which can be also written as
\be\label{GWField2}
\hat{h}_{ij}^{\rm TT}(\Omega,\B{x})=\frac{1}{2M_G}\mathbb{L}^{\rm TT}(\B{x})\left[\int \frac{d^3\B{k}}{(2\pi)^3}\frac{\hat{T}_{ij}(\Omega,\B{k})e^{i\B{k}\cdot\B{x}}}{\omega_k^2-\Omega^2}\right]
\ee
where $\hat{T}_{ij}(\Omega,\B{k})$ is the time and spatial Fourier transform of the stress energy tensor of the probe, and, neglecting terms of $O(q/L)$ and $O(v^2/c^2)$, is simply the stress energy tensor of the optical mode $\hat{T}^{EM}_{ij}$. It is given  by the 
\be\label{stressEnergy}
\hat{T}_{ij}(\Omega,\B{k})=\hat{T}^{EM}_{ij}(\Omega,\B{k})=\delta_{ix}\delta_{jx}\frac{\omega_0\bar{\alpha}}{c^2}\hat{\alpha}_1(\Omega)
\ee
The notation $\mathbb{L}^{\rm TT}(\B{x})$ is shorthand for the TT projection operator, which projects each Fourier component $\hat{T}_{ij}(\Omega,\B{k})$ to its TT components with respect to propagation vector $\B{k}$, and accounts for the presence of the polarization tensors in Eq.~\eqref{GWField}. Explicitly, this operation on some general tensor field $f_{ij}(\B{x})$ is given by
\begin{align}
\mathbb{L}^{\rm TT}(\B{x})\left[f_{ij}(\B{x})\right]
=&\int d^3\B{x}'\int d^3\B{x}''\,f_{lm}(\B{x}'')\notag\\
&\bigg[
\C{P}_{il}(\B{x},\B{x}{'})\C{P}_{jm}(\B{x}',{\B{x}}{''})\notag\\
&-\frac{1}{2}\C{P}_{ij}(\B{x},{\B{x}}{'})\C{P}_{mn}(\B{x}',\B{x}{''})\delta^n_l
\bigg]\label{TTProjector}
\end{align}
where $\C{P}_{ij}(\B{x},\B{x}')$ is the transverse projection operator for vector plane waves, such that $\int d^3\B{x}'\C{P}_{ij}(\B{x},\B{x}')v_je^{i\B{k}\cdot\B{x}'}=v^{\bot}_i e^{i\B{k}\cdot\B{x}}$, and is equal to 
\be
\C{P}_{ij}(\B{x},\B{x}')=\delta_{ij}\delta^3(\B{x}-\B{x}')-\partial_i G(\B{x},\B{x}')\partial_{j'}
\ee
where $G(\B{x},\B{x}')=-1/(4\pi|\B{x}-\B{x}'|)$ is the Green's function of the $\BS{\nabla}^2$ operator. The equivalence of Eq.~\eqref{GWField2} to Eqs.~\eqref{GWField} then follows readily from applying Eq.~\eqref{stressEnergy} and~\eqref{TTProjector}.

From Eq.~\eqref{GWField2} we may obtain the result from Einstein's field equations by identifying
\be
\hat{T}_{ij}(\Omega,\B{k})\approx\int d^3\B{x}\;\hat{T}_{ij}(\Omega,\B{x})
\ee
which holds under the slow-motion approximation and which states that the Fourier $\B{k}$-mode of the stress energy tensor is approximately equal to its own volume integral. Then, performing the inverse transform on Eq.~\eqref{GWField2} into the time domain in the far zone, we obtain form the equations of motion in Eqs.~\eqref{gravEOM} our final expression for the GW field:
\be\label{GWFieldFinal}
\hat{h}_{ij}^{\rm TT}(t,\B{x})=\frac{4G}{c^2}\mathbb{L}^{\rm TT}(\B{x})\left[
\frac{\int d^3\B{x}'\;\hat{T}_{ij}(t-|\B{x}|/c,\B{x}')}{|\B{x}|}
\right]
\ee
To see that this is equivalent to the quadrupole formula in Eq.~\eqref{quadrupoleFormula}, we point that that in the far zone approximation the argument of the TT projection operator $\mathbb{L}^{\rm TT}$ depends only on the distance $r=|\B{x}|$, and therefore Eq.~\eqref{TTProjector} reduces to a simple form where it can be written in terms of the transverse projection operator for radially traveling waves:
\be\label{RadialProjector}
\mathbb{L}^{\rm TT}\left[f_{ij}\right]=P_{il}f_{lm}P_{mj}-\frac{1}{2}P_{ij}{\rm Tr}\left\{\BS{P}\BS{f}\right\}
\ee
where $P_{lm}=\delta_{lm}-{x_lx_m}/{r^2}$. In this way, it follows that Eqs.~\eqref{GWFieldFinal} and~\eqref{quadrupoleFormula} are equivalent (with a final cosmetic step invoking stress-energy conservation \cite{MTW}), and therefore that our equations of motion are consistent with classical general relativity.

In summary, we have reproduced the quadrupole formula for gravitational-wave generation using a fully quantum formalism. This serves as a check to our theoretical framework, as well as a theoretical basis for discussing quantum GW states generated by quantum matter.

\section{Probe Dynamics}\label{probeDynamics}
Previously, we pointed out that in the presence of a large excitation the GW field can be decomposed into classical and quantum components, whereby the interaction Hamiltonian takes the form of Eq.~\eqref{HInt}. Correspondingly, the GW effects that appear in $\dot{\hat{\alpha}}_2$ consists of a predetermined classical component $\C{h}_s(t)$ as well as a quantum component involving $\hat{h}_{\lambda}(\B{k})$. In previous treatments one considered only $\C{h}_s(t)$, which acts as on the probe as a fixed external force \cite{kimble2001}. However, as discussed in section \ref{GWDynamics} and as evident in Eq.\eqref{GWField}, $\hat{h}_\lambda(\B{k})$ has contributions from the probe itself which is responsible for quantum coherent GW backaction. In this section, we obtain the equations of motion incorporating this quantum contribution and discuss their consequences.

\subsection{Solving Equations of Motion in the Laplace Domain}

Let us solve for $\hat{\alpha}_1,\,\hat{\alpha}_2$'s equations of motion in Eqs.~\eqref{ampEOM} and~\eqref{phaseEOM} in the Laplace domain, defining $t=0$ to be the time of the probe's initial state preparation and using the transform pair
\be\label{laplaceTransform}
\hat{O}(s)=\int_0^\infty dt\; \hat{O}(t)e^{-st},\quad
\hat{O}(t)=\frac{1}{2\pi i}\int_{- i\infty +\epsilon}^{i\infty +\epsilon} ds\;\hat{O}(s)e^{st}
\ee
Then, each $\B{k}$-mode of the GW field is given by
\begin{align}
\hat{h}_\lambda(s,\B{k})=&\,\frac{s}{s^2+\omega_k^2}\hat{h}_\lambda(0,\B{k})+\frac{1}{M_G}\frac{\hat{\Pi}_\lambda(0,\B{k})}{s^2+\omega_k^2}\notag\\
&+\frac{\omega_0\bar{\alpha}}{2}\frac{1}{M_G}J_\lambda^*(\B{k})\frac{\hat{\alpha}_1(s)}{s^2+\omega_k^2}\label{GWFieldLaplace}
\end{align}
The difference between Eq.~\eqref{GWFieldLaplace} and our previous result in Eq.~\eqref{GWField} is that the Laplace transform solutions accounts for the initial values of the field, whereas the Fourier transform solution does not. These initial values, given by the first two terms on the right-hand side of Eq.~\eqref{GWFieldLaplace}, represent the input quantum GW fluctuations which we collectively denote by $\C{\hat{h}}_{\rm in}$, while the last term is the part of the GW field generated by the probe that leads to backaction.

Using Eq.~\eqref{GWFieldLaplace} and also substituting $\hat{q}(s)={\omega_0\bar{\alpha}}\,{\hat{\alpha}_1}/{mLs^2}$ into Eq.~\eqref{phaseEOM} (ignoring the initial state of the test mass, which is irrelevant for the interaction dynamics), we solve Eqs.~\eqref{ampEOM} and~\eqref{phaseEOM} for the optical mode in terms of the input field quadratures $\hat{\alpha}_{1,2}^{\rm in},\,\hat{\alpha}_{1,2}^{\rm out}$, which are associated with the mode operators defined in Eq.~\eqref{inputOutputOperators}, and obtain
\begin{align}
\begin{bmatrix}
s+\gamma & \Delta\\
-(1+\xi_{\rm BA})\Delta & s+\gamma
\end{bmatrix}
&\begin{pmatrix}
\hat{\alpha}_1\\
\hat{\alpha}_2
\end{pmatrix}\notag\\
=
\sqrt{2\gamma}\,
\begin{pmatrix}
\hat{\alpha}^{\rm in}_{1}\\
\hat{\alpha}^{\rm in}_{2}
\end{pmatrix}
&+\frac{\omega_0\bar{\alpha}}{2}
\begin{pmatrix}
0\\
1
\end{pmatrix}
\left[\C{h}_s+\hat{\C{h}}_{\rm in}\right]\label{cavityEOM}
\end{align}
where $\xi_{\rm BA}$ is the modification to optical mode response due to backaction from the test mass and GW field, and is given by
\be\label{responseMod}
\xi_{\rm BA}(s)=\frac{1}{\Delta}\left(\frac{\epsilon_{q}}{s^2}+\epsilon_{GW}s\right) 
\ee
with
\begin{subequations}
\begin{align}
\epsilon_q & =\frac{(\omega_0\bar{\alpha})^2}{mL^2} \\
\epsilon_{GW}&=\frac{8G}{15c^5}(\omega_0\bar{\alpha})^2
\end{align}
\end{subequations}
and where we have included the effect of the predetermined classical component of the field $\C{h}_s$ in our solution. Here $\epsilon_q$ arises due to interaction between light and test mass, while $\epsilon_{GW}$ arises due to interaction between light and the gravitational field.

We remark that for GW backaction we have only considered effects due to outgoing gravitational waves. The leading order Feynman diagrams for back action processes involve a graviton propagator between in and out matter states, which in principle should include contributions from longitudinal and timelike gravitons if we were to account for all gravitational effects. Restricting our attention to the TT modes ignores time-symmetric self-gravity effects such as the Newtonian self-potential, but those are well separated from the leading order time-asymmetric term that is the GW backaction.
\subsection{Tuned Configuration}
At times long enough such that the initial state of the probe is forgotten, for the tuned configuration with $\Delta=0$ we obtain the cavity mode solutions:
\begin{subequations}
\be\label{ampSolnTuned}
\hat{\alpha}_1=\frac{\sqrt{2\gamma}}{s+\gamma}\hat{\alpha}^{\rm in}_{1}
\ee
\be\label{phaseSolnTuned}
\hat{\alpha}_2=\frac{\sqrt{2\gamma}}{s+\gamma}\hat{\alpha}^{\rm in}_{2}+
\xi_{\rm BA}\frac{\sqrt{2\gamma}}{(s+\gamma)^2}\hat{\alpha}^{\rm in}_{1}
+\frac{\omega_0\bar{\alpha}}{2}\left(\C{h}_s+\hat{\C{h}}_{\rm in}\right)
\ee
\end{subequations}
Inspecting the above equations we see that in that the backaction term does not modify the dynamical response of either quadrature. Instead, it introduces additional fluctuations to the phase quadrature $\hat\alpha_2$ which modifies the shape of the output noise ellipse. This is shown explicitly in the noise input-output relations given by 
\begin{equation}
\hat{\alpha}^{\rm out}_j=\hat{\alpha}^{\rm in}_{j} -\sqrt{2\gamma}\hat{\alpha}_j
\end{equation}
for $j=1,2$, and the input-output relations themselves can be derived from the EOM for the external pump field using $H_{\rm ext}$. Substituting in the solutions for $\hat{\alpha}_j$ given in Eqs.~\eqref{ampSolnTuned} and~\eqref{phaseSolnTuned}, we find the fluctuating parts of the out-going quadratures
\begin{subequations}
\be\label{inOutAmp}
\hat\alpha^{\rm out}_1=e^{2i\beta}\hat{\alpha}^{\rm in}_{1}
\ee
\begin{align}
\hat{\alpha}^{\rm out}_{2}=e^{2i\beta}\left[\hat{\alpha}^{\rm in}_{2}-\left(\C{K}_{pd}+i\C{K}_{GW} \right)\hat{\alpha}^{\rm in}_{1}\right]-\sqrt{\frac{\gamma}{2}}\frac{\omega_0\bar{\alpha}\,\hat{\C{h}}_{\rm in}}{(s+\gamma)}\label{inOutPhase}
\end{align}
\end{subequations}
where $\beta$ is an uninteresting overall phase factor and $\C{K}_{pd}$ and $\C{K}_{GW}$ are the backaction terms due to the test mass and GWs respectively. Making the identification $s\rightarrow -i\Omega$, we have $\beta=\arctan({\Omega/\gamma})+\pi/2$ and
\begin{subequations}\label{backActionPars}
\be
\C{K}_{pd}(\Omega)= \frac{1}{m}\left(\frac{\omega_0\bar{\alpha}}{L}\right)^2\frac{2\gamma}{\Omega^2(\gamma^2+\Omega^2)}
\ee
\be
\C{K}_{GW}(\Omega)=\frac{2\Omega\gamma\epsilon_{GW}}{\gamma^2+\Omega^2}
\ee
\end{subequations}
 
On the right-hand side of Eq.~\eqref{inOutPhase}, the term that contains $\C{K}_{pd}$ gives rise to the well known ponderomotive effect which causes a rotation and squeezing of the input noise ellipse, with rotation angle $\theta=\arctan(\C{K}_{pd}/2)$, squeeze angle $\phi={\rm arccot}(\C{K}_{pd}/2)/2$, and squeeze factor $r={\rm arcsinh}(\C{K}_{pd}/2)$~\cite{kimble2001}. On the other hand, the term containing $\mathcal{K}_{\rm GW}$, which gives rise to GW backaction, is imaginary, and therefore must be associated with additional fluctuations in $\hat{\alpha}^{\rm out}_{2}$ in order for $\left[\hat{\alpha}^{\rm out}_{2}(t),\hat{\alpha}^{\rm out}_{2}(t')\right]=0$. Since $\hat\alpha^{\rm out}_{1,2}(t)$ are out-going fields at different moments of time, and therefore are independent degrees of freedom, this commutation relation must be satisfied.

Indeed, we see in Eq.~\eqref{inOutPhase} that there are fluctuations from the GW field which ensure that the output phase quadrature commute at different times, and additionally enlarges the total area of the output noise ellipse. Unfortunately, these additional fluctuations cannot be removed without access to quantum GW degrees of freedom and will therefore introduce additional noise, albeit at $O(\epsilon_{GW}^2)$.
\subsection{Detuned Configuration}
When the optical drive is detuned from cavity resonance, the amplitude and phase quadratures of the cavity mode rotate into each other. In the presence of backaction, this results in modifications to their dynamical response functions, in contrast to the tuned configuration. For simplicity, we ignore the ponderomotive backaction due to the test mass by taking $m\rightarrow \infty$ and focus solely on gravitational effects. Then, solving Eq.~\eqref{cavityEOM} for $\Delta\neq 0$ we find
\begin{subequations}
\be\label{ampSolnDetuned}
\hat{\alpha}_1=\sqrt{2\gamma}\left[\chi_1\hat{\alpha}^{\rm in}_{1}-\chi_2\hat{\alpha}^{\rm in}_{2}\right]-\frac{\omega_0\bar{\alpha}}{2}\chi_2\left(\C{h}_s+\hat{\C{h}}_{\rm in}\right)
\ee
\begin{align}
\hat{\alpha}_2=\sqrt{2\gamma}\bigg[\left(1+\frac{\epsilon_{GW}s}{\Delta}\right)&\chi_2\hat{\alpha}^{\rm in}_{1}+\chi_1\hat{\alpha}^{\rm in}_{2}\bigg]\notag\\
&+\frac{\omega_0\bar{\alpha}}{2}\chi_1\left(\C{h}_s+\hat{\C{h}}_{\rm in}\right)\label{phaseSolnDetuned}
\end{align}
\end{subequations}
for the response functions
\begin{subequations}\label{response}
\be
\chi_1=\frac{s+\gamma}{(s+\gamma)^2+\Delta^2+\epsilon_{GW}\Delta s}
\ee
\be
\chi_s=\frac{\Delta}{(s+\gamma)^2+\Delta^2+\epsilon_{GW}\Delta s}
\ee
\end{subequations}
The optical mode's response to external drive is modified through $\epsilon_{GW}$, which can be encapsulated by shifts in the effective damping and detuning of the optical mode:
\be\label{effectiveRates}
\tilde{\gamma}=\gamma+\frac{\epsilon_{GW}\Delta}{2},\quad
\tilde{\Delta}=\Delta-\frac{\epsilon_{GW}\gamma}{2}
\ee

Since $\Delta$ can take positive or negative values, the backaction can either augment or reduce the effective cavity damping rate $\gamma$. This is due to the amplitude quadrature being coherently added to itself as the amplitude and phase quadratures rotate into each other through the $\Delta \hat{a}_2$ term in its own EOM. Depending the the sign of $\Delta$, it can either beat destructively ($\Delta>0$ or red-detuned), or constructively ($\Delta<0$ or blue-detuned).
\begin{figure}[h]\label{backaction}
\includegraphics[scale=0.4]{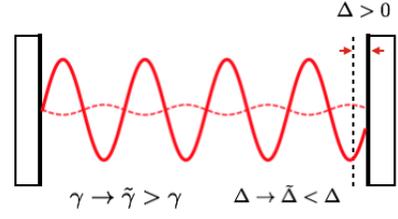}
\caption{Illustration of quantum coherent backaction effects onto the cavity mode due to GW interaction in the presence of detuning. The GWs generated by the $\hat{\alpha}_1$ acts back on $\hat{\alpha}_2$ in such a way that causes the field to beat coherently with itself. The above shows the case for red-detuning where $\Delta>0$. The solid red line represents the cavity mode in the absence of backaction, while the dotted red line represents the contribution due to GW backaction. This effect is quantified by changes to the cavity's effective damping and detuning rate so that $\tilde{\gamma}= \gamma+\epsilon_{GW}\Delta/2$ and $\tilde{\Delta}=\Delta -\epsilon_{GW}\gamma/2$.}
\end{figure}

\subsection{Backaction in the Newtonian Gauge}
To understand the GW backaction more intuitively, we now go to the Newtonian gauge and show that it is in fact the quantum analogue of the radiation reaction potential. The radiation reaction potential, or $\Phi_{\rm react}$, is the leading order time-asymmetric GR correction to the Newtonian potential, and derives from the outgoing GWs radiated by the probe's time dependent mass quadrupole moment. The correction accounts for the consequent loss of energy which leads to damping of the motion, an effect known as radiative damping \cite{MTW}. To see this explicitly, we write the EOM of the test mass in the Newtonian gauge under radiation pressure and the radiation reaction force:
\be
\ddot{\hat{q}}=-\frac{\partial }{\partial x}\Phi_{\rm react} + \frac{\omega_0\bar{\alpha}}{mL}\hat{\alpha}_1,\quad
\Phi_{\rm react}(\B{x})=\frac{G}{5c^5}\C{I}^{(5)}_{jk}x^j x^k
\ee
where $\mathcal{I}_{jk}$ is the reduced mass quadrupole moment tensor, and the superscript represents the number of time derivatives. We find that the reaction force on the system evaluates to 
\begin{equation}
\label{eqFreact}
F_{\rm react}=-\frac{8 G}{15 c^5}mL^2\hat{q}^{(5)}
\end{equation}
Since the reaction force is dependent on $\hat{q}$ itself, it serves to modify the response of $\hat{q}$ to radiation pressure force. Here we are interested only in the backaction effects of GW interaction and have therefore suppressed the input GW field. Assuming that the probe is operating under steady state, we may solve the probe EOM in the Fourier domain
\be\label{NTTestMass}
\hat{q}(\Omega)=\left[\chi_q^{(0)}+\delta\chi_q\right]\frac{\omega_0\bar{\alpha}}{L}\hat{\alpha}_1(\Omega)
\ee
for free mass susceptibility $\chi_q^{(0)}$ and its perturbation $\delta\chi_q$
\be\label{massResponse}
\chi_q^{(0)}=-\frac{1}{m\Omega^2},\quad
\delta\chi_q=-i\frac{8G}{15c^5}mL^2\Omega^3\chi_q^{(0)}
\ee

In the Newtonian gauge the optical mode does not interact gravitationally, and the Heisenberg EOM for its quadratures depends only on the input optical field and the test mass dynamics, which in the tuned configuration is given by $\hat{\alpha}_1(\Omega)(\gamma-i\Omega)=\sqrt{2\gamma}\hat{\alpha}^{\rm in}_{1}(\Omega )$ and $\hat{\alpha}_2(\Omega)(\gamma-i\Omega)=({\omega_0\bar{\alpha}}/{L})\hat{q}(\Omega)+\sqrt{2\gamma}\hat{\alpha}^{\rm in}_{2}(\Omega)$. Substituting Eq.~\eqref{NTTestMass} into the expression for $\hat{\alpha}_2$, we find
\begin{subequations}
\be\label{NTcavityAmp}
\hat{\alpha}_1=\frac{\sqrt{2\gamma}}{\gamma-i\Omega}\hat{\alpha}^{\rm in}_{1}
\ee
\be\label{NTcavityPhase}
\hat{\alpha}_2=\frac{\omega_0^2\bar{\alpha}^2}{L^2}\left(\chi_q^{(0)}+\delta\chi_q\right)\frac{\hat{\alpha}_1}{\gamma-i\Omega}+\frac{\sqrt{2\gamma}}{\gamma-i\Omega}\hat{\alpha}_{in}^{(2)}
\ee
\end{subequations}
where $\chi_q^{(0)}=-(m\Omega^2)^{-1}$ is the free test mass response and $\delta\chi_q$ is the GW correction. Making the substitution $-i\Omega\rightarrow s$, we find that modification to test mass response deriving from $\Phi_{\rm react}$ exactly corresponds to the GW backaction term in Eq.~\eqref{cavityEOM} for $\Delta\rightarrow 0$
\be
\left(\frac{\omega_0\bar{\alpha}}{L}\right)^2\frac{\delta\chi_q}{\gamma-i\Omega}\rightarrow \frac{\epsilon_{GW}s}{s+\gamma}
\ee
From here it follows that identical input-output relations to Eq.s~\eqref{inOutAmp} and~\eqref{inOutPhase} may be obtained. Here we note that the fifth order time derivative in Eq.~\eqref{eqFreact} makes the radiation reaction force time asymmetric, which results in $\mathcal{K}_{\rm GW}$ being imaginary.

It is straightforward to see that an analysis for the detuned configuration would yield a similar result, since the only difference in the EOMs for $\hat{\alpha}_1$ and $\hat{\alpha}_2$ between the tuned and detuned cases in the Newtonian gauge would be the addition of the terms $-\Delta\hat{\alpha}_2$ to the RHS of Eq.~\eqref{NTcavityAmp} and $\Delta \hat{\alpha}_1$ to that of Eq.~\eqref{NTcavityPhase}. This yields identical expressions to Eq.~\eqref{cavityEOM}.

We have confirmed that in both gauges the same backaction effect appears in the output field which is being measured, as must be the case. However, there are some interesting points that arises from comparing the different interpretations offered by each gauge. First, in the Newtonian gauge, the underlying physical mechanism for the backaction is the modification of test mass response to radiation pressure due to the radiative damping. Thus, the GW backaction can be interpreted as a correction to the response function of the test mass to external forces (i.e. radiation pressure), given by $\delta \chi_q$ in Eq.~\eqref{massResponse}. Correspondingly, $\C{K}_{GW}$ can be interpreted as a correction to the ponderomotive backaction $\C{K}_{pd}$ in Eqs.~\eqref{backActionPars}. Viewed in this way, one might intuitively expect that in the limit $m\rightarrow\infty$ the infinitely massive mirror would be unperturbed by radiation pressure, and therefore both $\C{K}_{pd}$ and its correction $\C{K}_{GW}$ should go to zero. However, as demonstrated in the TT gauge, the GW backaction appears without consideration of test mass dynamics, and without regard to its mass. A more careful look at the Newtonian gauge reveals that this result is consistent, due to the fact that $F_{\rm react}$ is mass dependent, and therefore, under weak equivalence, the inertial mass in $\chi_q^{(0)}$ which appears in the denominator of $\delta\chi_q$ cancels the gravitational mass in $F_{\rm react}$ which appears in its numerator. The gravitational radiation reaction therefore ensures that even an infinitely massive object will respond to external forces.

The second point of contrast is that in the Newtonian gauge, the motion of the test mass is indeed damped due to energy loss though GW radiation. However, in the TT gauge, neither the test mass nor the optical mode experiences damping. Again, this seemingly paradoxical observation may be resolved when one recognizes that the energy that drives the test mass motion is from the optical pump, which also provides the energy radiated away in GWs. This is clear from Eqs.~\eqref{GWFieldFinal} and ~\eqref{cavityEOM}, where the generation of GWs is shown to depend on $T_{ij}\sim \hat{\alpha}_1$ which depends only on the input optical field, as shown in Eq.~\eqref{NTcavityAmp}. That the energy lost to GW radiation is sourced from the external optical pump also addresses the counterintuitive result that GW backaction can reduce the effective cavity damping rate, since the damping of the cavity mode does not directly correspond to energy exchange with the GW field.

\subsection{Role of Quantum GW Fluctuations in Backaction}
Although radiation reaction can be traced back to the classical radiation reaction potential, in order to consistently include its effects on quantum matter, one must necessarily quantize the GW fluctuations. This is clear from Eqs.~\eqref{ampSolnDetuned} and~\eqref{phaseSolnDetuned}, where including the $\epsilon_{GW}$ backaction effect without also including the $\hat{\C{h}}_{\rm in}$ fluctuations will result in a modified commutator between the cavity mode quadratures
\be\label{wrongCommutator}
\left[\hat{\alpha}_1(t),\hat{\alpha}_2(t)\right]=i\hbar\left[1-\frac{\epsilon_{GW}\Delta}{2\gamma}\left(1-e^{2\tilde{\gamma}t}\right)\right]
\ee
and only by also including $\hat{\C{h}}_{\rm in}$ will we recover the canonical commutation relation $\left[\hat{\alpha}_1(t),\hat{\alpha}_2(t)\right]=i\hbar$. This result furthers our understanding of how gravity interacts with quantum matter and advances the point of view that gravity must ultimately be quantum, since the predictions of semiclassical gravity upon including the radiation damping correction will result in Eq~\eqref{wrongCommutator}, in violation of the uncertainty principle.

\section{Conclusion}
To summarize, in this work we developed a framework for the Hamiltonian formulation of the interaction between LIGO or a similar laser interferometer with gravity, such that both the GW field and the matter probe are dynamical degrees of freedom of the total system. To do so, we needed to address the issues of i) gauge fixing for the gravitational field and ii) formulating the optomechanical interaction using the action principle when the equations of motion are unknown. With regard to the former, we have argued that keeping only leading order interactions between the detector and gravitational waves allows us to fix the action in the TT gauge. For the latter, we have shown how to derive the optomechanical interaction from the action of an electromagnetic field with a dynamical boundary condition. Our completed formulation allows for the leading order quantum treatment of the interactions between LIGO and gravitational waves, for which, in the limit of classical gravity, recovers known equations of motion for LIGO's output fields as well as the quadrupole moment formula for GW generation. This serves as a verification for the framework, which, allowing for quantum gravity, additionally predicts quantum coherent backaction effects that correspond to the classical radiation reaction potential in general relativity. We have shown that in order for this well studied potential to be consistently applied to quantum matter such that commutation relations are preserved, the gravitational field itself must be quantum. This suggests that testing the effects of gravity at post-Newtonian orders may offer some insight into the quantum versus classical nature of gravity.

Even in the classical limit for gravity, our formulation presents an alternative interpretation of LIGO's detections. In the Newtonian gauge used by previous analyses, the interaction occurs between the test mass and the signal which subsequently modulates the cavity mode; in the TT gauge, the interaction occurs directly with the cavity mode. While the signal appears in the same way in the output fields that we measure, the latter point of view provides a rigorous foundation for calculating the fundamental limit to LIGO's detection sensitivity in the form of its qCRB. The qCRB follows readily from the Hamiltonian in Eq.~\eqref{HInt}, and can be expressed in the form of a minimum bound on the noise spectral density for the estimation error on $\C{h}_s(t)$, or
\be
S_{\Delta\C{h}_s}(\Omega)\geq \frac{\hbar^2}{(\omega_0\bar{\alpha})^2S_{\alpha_1}(\Omega)}
\ee
where $\Delta\C{h}_s(t)$ is the residual between the signal and our estimate, and the noise spectral density for any stationary process $x(t)$ is given by $S_x(\Omega)=\int_{-\infty}^{\infty} d\tau\;e^{i\Omega \tau}\langle x(t+\tau)x(t)\rangle$. We also point out that we have not considered the decohering effects of GW fluctuations, which again follows from the formalism. In fact, it can be shown that the qCRB, decoherence, and radiation are fundamentally and quantifiably related, and we therefore delay a more detailed discussion of these processes and their relations to an accompanying paper. Finally, we remark that our framework offers a theoretical foundation consistent with current observations that can be extended to study alternative theories of gravity in LIGO in the form of modifications to the Einstein-Hilbert action. It provides a basic model for investigating the quantum versus classical nature of GR or modified gravity and how their features manifest themselves in quantum measurement.

\section{Acknowledgement} Research of Y.C.\ and B.P.\ are supported by the NSF Grants PHY-1708212, PHY-1404569 and  PHY-1708213 and the Simons Foundation. We thank H.\ Miao and Y.\ Ma for their early contributions to this work, and also thank them along with R. \ Adhikari for productive discussions. 

\bibliographystyle{apsrev}
\bibliography{gravityRefs}

\begin{thebibliography}{32}
\expandafter\ifx\csname natexlab\endcsname\relax\def\natexlab#1{#1}\fi
\expandafter\ifx\csname bibnamefont\endcsname\relax
  \def\bibnamefont#1{#1}\fi
\expandafter\ifx\csname bibfnamefont\endcsname\relax
  \def\bibfnamefont#1{#1}\fi
\expandafter\ifx\csname citenamefont\endcsname\relax
  \def\citenamefont#1{#1}\fi
\expandafter\ifx\csname url\endcsname\relax
  \def\url#1{\texttt{#1}}\fi
\expandafter\ifx\csname urlprefix\endcsname\relax\def\urlprefix{URL }\fi
\providecommand{\bibinfo}[2]{#2}
\providecommand{\eprint}[2][]{\url{#2}}

\bibitem[{\citenamefont{Abbott et~al.}(2016)}]{GW150914}
\bibinfo{author}{\bibfnamefont{B.~P.} \bibnamefont{Abbott}}
  \bibnamefont{et~al.} (\bibinfo{collaboration}{LIGO Scientific Collaboration,
  Virgo Collaboration}), \bibinfo{journal}{Phys. Rev. Lett.}
  \textbf{\bibinfo{volume}{116}} (\bibinfo{year}{2016}).

\bibitem[{\citenamefont{Acernese et~al.}(2018)}]{virgo}
\bibinfo{author}{\bibfnamefont{F.}~\bibnamefont{Acernese}} \bibnamefont{et~al.}
  (\bibinfo{collaboration}{Virgo Collaboration}),
  \bibinfo{journal}{arXiv:1807.03275}  (\bibinfo{year}{2018}).

\bibitem[{\citenamefont{Aso et~al.}(2013)}]{kagra}
\bibinfo{author}{\bibfnamefont{Y.}~\bibnamefont{Aso}} \bibnamefont{et~al.}
  (\bibinfo{collaboration}{KAGRA Collaboration}), \bibinfo{journal}{Phys. Rev.
  D} \textbf{\bibinfo{volume}{88}} (\bibinfo{year}{2013}).

\bibitem[{\citenamefont{Abbott et~al.}(2017{\natexlab{a}})}]{GW170814}
\bibinfo{author}{\bibfnamefont{B.~P.} \bibnamefont{Abbott}}
  \bibnamefont{et~al.} (\bibinfo{collaboration}{LIGO Scientific Collaboration,
  Virgo Collaboration}), \bibinfo{journal}{Phys. Rev. Lett.}
  \textbf{\bibinfo{volume}{119}} (\bibinfo{year}{2017}{\natexlab{a}}).

\bibitem[{\citenamefont{{The LIGO Scientific Collaboration and the Virgo
  Collaboration}}(2018)}]{nontensorialGWpulsar}
\bibinfo{author}{\bibnamefont{{The LIGO Scientific Collaboration and the Virgo
  Collaboration}}}, \bibinfo{journal}{Phys. Rev. Lett.}
  \textbf{\bibinfo{volume}{120}} (\bibinfo{year}{2018}).

\bibitem[{\citenamefont{Yunes et~al.}(2016)\citenamefont{Yunes, Yagi, and
  Pretorius}}]{yunes2016}
\bibinfo{author}{\bibfnamefont{N.}~\bibnamefont{Yunes}},
  \bibinfo{author}{\bibfnamefont{K.}~\bibnamefont{Yagi}}, \bibnamefont{and}
  \bibinfo{author}{\bibfnamefont{F.}~\bibnamefont{Pretorius}},
  \bibinfo{journal}{Phys. Rev. D} \textbf{\bibinfo{volume}{94}}
  (\bibinfo{year}{2016}).

\bibitem[{\citenamefont{Amelino-Camelia}(1999)}]{camelia1999}
\bibinfo{author}{\bibfnamefont{G.}~\bibnamefont{Amelino-Camelia}},
  \bibinfo{journal}{Nature} \textbf{\bibinfo{volume}{398}}
  (\bibinfo{year}{1999}).

\bibitem[{\citenamefont{Abbott et~al.}(2017{\natexlab{b}})}]{GW170104}
\bibinfo{author}{\bibfnamefont{B.~P.} \bibnamefont{Abbott}}
  \bibnamefont{et~al.} (\bibinfo{collaboration}{LIGO Scientific Collaboration,
  Virgo Collaboration}), \bibinfo{journal}{Phys. Rev. Lett.}
  \textbf{\bibinfo{volume}{118}} (\bibinfo{year}{2017}{\natexlab{b}}).

\bibitem[{\citenamefont{Arun and Will}(2009)}]{arun2009}
\bibinfo{author}{\bibfnamefont{K.~G.} \bibnamefont{Arun}} \bibnamefont{and}
  \bibinfo{author}{\bibfnamefont{C.~M.} \bibnamefont{Will}},
  \bibinfo{journal}{Class. Quantum Grav.} \textbf{\bibinfo{volume}{26}}
  (\bibinfo{year}{2009}).

\bibitem[{\citenamefont{Miao et~al.}(2017)\citenamefont{Miao, Adhikari, Ma,
  Pang, and Yanbei}}]{miao2017}
\bibinfo{author}{\bibfnamefont{H.}~\bibnamefont{Miao}},
  \bibinfo{author}{\bibfnamefont{R.~X.} \bibnamefont{Adhikari}},
  \bibinfo{author}{\bibfnamefont{Y.}~\bibnamefont{Ma}},
  \bibinfo{author}{\bibfnamefont{B.}~\bibnamefont{Pang}}, \bibnamefont{and}
  \bibinfo{author}{\bibfnamefont{C.}~\bibnamefont{Yanbei}},
  \bibinfo{journal}{Phys. Rev. Lett.} \textbf{\bibinfo{volume}{119}}
  (\bibinfo{year}{2017}).

\bibitem[{\citenamefont{Downes et~al.}(2017)\citenamefont{Downes, van Meter,
  Knill, Milburn, and Caves}}]{downes2017}
\bibinfo{author}{\bibfnamefont{T.~G.} \bibnamefont{Downes}},
  \bibinfo{author}{\bibfnamefont{J.~R.} \bibnamefont{van Meter}},
  \bibinfo{author}{\bibfnamefont{E.}~\bibnamefont{Knill}},
  \bibinfo{author}{\bibfnamefont{G.~J.} \bibnamefont{Milburn}},
  \bibnamefont{and} \bibinfo{author}{\bibfnamefont{C.~M.} \bibnamefont{Caves}},
  \bibinfo{journal}{Phys. Rev. D} \textbf{\bibinfo{volume}{96}}
  (\bibinfo{year}{2017}).

\bibitem[{\citenamefont{Kafri et~al.}(2014)\citenamefont{Kafri, Taylor, and
  Milburn}}]{kafri2014}
\bibinfo{author}{\bibfnamefont{D.}~\bibnamefont{Kafri}},
  \bibinfo{author}{\bibfnamefont{J.~M.} \bibnamefont{Taylor}},
  \bibnamefont{and} \bibinfo{author}{\bibfnamefont{G.~J.}
  \bibnamefont{Milburn}}, \bibinfo{journal}{New J. Phys.}
  \textbf{\bibinfo{volume}{16}} (\bibinfo{year}{2014}).

\bibitem[{\citenamefont{Kleckner et~al.}(2016)\citenamefont{Kleckner, Pikovski,
  Jeffrey, Ament, Eliel, van~den Brink, and Bouwmeester}}]{kleckner2008}
\bibinfo{author}{\bibfnamefont{D.}~\bibnamefont{Kleckner}},
  \bibinfo{author}{\bibfnamefont{I.}~\bibnamefont{Pikovski}},
  \bibinfo{author}{\bibfnamefont{E.}~\bibnamefont{Jeffrey}},
  \bibinfo{author}{\bibfnamefont{L.}~\bibnamefont{Ament}},
  \bibinfo{author}{\bibfnamefont{E.}~\bibnamefont{Eliel}},
  \bibinfo{author}{\bibfnamefont{J.}~\bibnamefont{van~den Brink}},
  \bibnamefont{and}
  \bibinfo{author}{\bibfnamefont{D.}~\bibnamefont{Bouwmeester}},
  \bibinfo{journal}{Phys. Rev. Lett.} \textbf{\bibinfo{volume}{116}}
  (\bibinfo{year}{2016}).

\bibitem[{\citenamefont{Yang et~al.}(2013)\citenamefont{Yang, Miao, Lee, Helou,
  and Chen}}]{yang2013}
\bibinfo{author}{\bibfnamefont{H.}~\bibnamefont{Yang}},
  \bibinfo{author}{\bibfnamefont{H.}~\bibnamefont{Miao}},
  \bibinfo{author}{\bibfnamefont{D.-S.} \bibnamefont{Lee}},
  \bibinfo{author}{\bibfnamefont{B.}~\bibnamefont{Helou}}, \bibnamefont{and}
  \bibinfo{author}{\bibfnamefont{Y.}~\bibnamefont{Chen}},
  \bibinfo{journal}{Phys. Rev. Lett.} \textbf{\bibinfo{volume}{110}}
  (\bibinfo{year}{2013}).

\bibitem[{\citenamefont{Pikovski et~al.}(2013)\citenamefont{Pikovski, Vanner,
  Aspelmeyer, Kim, and Brukner}}]{pikovski2012}
\bibinfo{author}{\bibfnamefont{I.}~\bibnamefont{Pikovski}},
  \bibinfo{author}{\bibfnamefont{M.~R.} \bibnamefont{Vanner}},
  \bibinfo{author}{\bibfnamefont{M.}~\bibnamefont{Aspelmeyer}},
  \bibinfo{author}{\bibfnamefont{M.~S.} \bibnamefont{Kim}}, \bibnamefont{and}
  \bibinfo{author}{\bibfnamefont{{\u C}.}~\bibnamefont{Brukner}},
  \bibinfo{journal}{Nat} \textbf{\bibinfo{volume}{110}} (\bibinfo{year}{2013}).

\bibitem[{\citenamefont{Belenchia et~al.}(2016)\citenamefont{Belenchia,
  Benincasa, Liberati, Marin, Marino, and Ortolan}}]{belenchia2016}
\bibinfo{author}{\bibfnamefont{A.}~\bibnamefont{Belenchia}},
  \bibinfo{author}{\bibfnamefont{D.~M.~T.} \bibnamefont{Benincasa}},
  \bibinfo{author}{\bibfnamefont{S.}~\bibnamefont{Liberati}},
  \bibinfo{author}{\bibfnamefont{F.}~\bibnamefont{Marin}},
  \bibinfo{author}{\bibfnamefont{F.}~\bibnamefont{Marino}}, \bibnamefont{and}
  \bibinfo{author}{\bibfnamefont{A.}~\bibnamefont{Ortolan}},
  \bibinfo{journal}{Phys. Rev. Lett.} \textbf{\bibinfo{volume}{116}}
  (\bibinfo{year}{2016}).

\bibitem[{\citenamefont{Anastopoulos and Hu}(2012)}]{anastopoulos2013}
\bibinfo{author}{\bibfnamefont{C.}~\bibnamefont{Anastopoulos}}
  \bibnamefont{and} \bibinfo{author}{\bibfnamefont{B.~L.} \bibnamefont{Hu}},
  \bibinfo{journal}{Class. Quantum Grav.} \textbf{\bibinfo{volume}{30}}
  (\bibinfo{year}{2012}).

\bibitem[{\citenamefont{Blencowe}(2013)}]{blencowe2013}
\bibinfo{author}{\bibfnamefont{M.~P.} \bibnamefont{Blencowe}},
  \bibinfo{journal}{Phys. Rev. Lett.} \textbf{\bibinfo{volume}{111}}
  (\bibinfo{year}{2013}).

\bibitem[{\citenamefont{Oniga and Wang}(2016{\natexlab{a}})}]{oniga2016}
\bibinfo{author}{\bibfnamefont{T.}~\bibnamefont{Oniga}} \bibnamefont{and}
  \bibinfo{author}{\bibfnamefont{C.~H.-T.} \bibnamefont{Wang}},
  \bibinfo{journal}{Phys. Rev. D} \textbf{\bibinfo{volume}{93}}
  (\bibinfo{year}{2016}{\natexlab{a}}).

\bibitem[{\citenamefont{Oniga and Wang}(2016{\natexlab{b}})}]{oniga2016bound}
\bibinfo{author}{\bibfnamefont{T.}~\bibnamefont{Oniga}} \bibnamefont{and}
  \bibinfo{author}{\bibfnamefont{C.~H.-T.} \bibnamefont{Wang}},
  \bibinfo{journal}{J. Phys. Conf. Ser.} \textbf{\bibinfo{volume}{845}}
  (\bibinfo{year}{2016}{\natexlab{b}}).

\bibitem[{\citenamefont{Damour and Esposito-Farese}(1992)}]{damour1992}
\bibinfo{author}{\bibfnamefont{T.}~\bibnamefont{Damour}} \bibnamefont{and}
  \bibinfo{author}{\bibfnamefont{G.}~\bibnamefont{Esposito-Farese}},
  \bibinfo{journal}{Class. Quantum Grav.} \textbf{\bibinfo{volume}{9}}
  (\bibinfo{year}{1992}).

\bibitem[{\citenamefont{Kimble et~al.}(2001)\citenamefont{Kimble, Levin,
  Matsko, Thorne, and Vyatchanin}}]{kimble2001}
\bibinfo{author}{\bibfnamefont{H.~J.} \bibnamefont{Kimble}},
  \bibinfo{author}{\bibfnamefont{Y.}~\bibnamefont{Levin}},
  \bibinfo{author}{\bibfnamefont{A.~B.} \bibnamefont{Matsko}},
  \bibinfo{author}{\bibfnamefont{K.~S.} \bibnamefont{Thorne}},
  \bibnamefont{and} \bibinfo{author}{\bibfnamefont{S.~P.}
  \bibnamefont{Vyatchanin}}, \bibinfo{journal}{Phys. Rev. D}
  \textbf{\bibinfo{volume}{65}} (\bibinfo{year}{2001}).

\bibitem[{\citenamefont{Misner et~al.}(1973)\citenamefont{Misner, Thorne, and
  Wheeler}}]{MTW}
\bibinfo{author}{\bibfnamefont{C.~W.} \bibnamefont{Misner}},
  \bibinfo{author}{\bibfnamefont{K.~S.} \bibnamefont{Thorne}},
  \bibnamefont{and} \bibinfo{author}{\bibfnamefont{J.~A.}
  \bibnamefont{Wheeler}}, \emph{\bibinfo{title}{Gravitation}}
  (\bibinfo{publisher}{W. H. Freeman and Company}, \bibinfo{year}{1973}).

\bibitem[{\citenamefont{Helstrom}(1969)}]{helstromQuantumEstimation}
\bibinfo{author}{\bibfnamefont{C.~W.} \bibnamefont{Helstrom}},
  \bibinfo{journal}{Journal of Statistical Physics}
  \textbf{\bibinfo{volume}{1}}, \bibinfo{pages}{231} (\bibinfo{year}{1969}).

\bibitem[{\citenamefont{Buonanno and Chen}(2003)}]{buonanno2003}
\bibinfo{author}{\bibfnamefont{A.}~\bibnamefont{Buonanno}} \bibnamefont{and}
  \bibinfo{author}{\bibfnamefont{Y.}~\bibnamefont{Chen}},
  \bibinfo{journal}{Phys. Rev. D} \textbf{\bibinfo{volume}{67}}
  (\bibinfo{year}{2003}).

\bibitem[{\citenamefont{Chen}(2013)}]{chen2013}
\bibinfo{author}{\bibfnamefont{Y.}~\bibnamefont{Chen}}, \bibinfo{journal}{J.
  Phys. B: At. Mol. Opt. Phys.} \textbf{\bibinfo{volume}{46}}
  (\bibinfo{year}{2013}).

\bibitem[{\citenamefont{Kiefer}(2012)}]{kiefer2012}
\bibinfo{author}{\bibfnamefont{C.}~\bibnamefont{Kiefer}},
  \emph{\bibinfo{title}{Quantum Gravity}} (\bibinfo{publisher}{Oxford
  University Press}, \bibinfo{year}{2012}).

\bibitem[{\citenamefont{Dirac}(1950)}]{diracConstrainedHamiltonian}
\bibinfo{author}{\bibfnamefont{P.~A.~M.} \bibnamefont{Dirac}},
  \bibinfo{journal}{Can. J. Math} \textbf{\bibinfo{volume}{2}}
  (\bibinfo{year}{1950}).

\bibitem[{\citenamefont{Gupta}(1952)}]{guptaLinearGrav}
\bibinfo{author}{\bibfnamefont{S.~N.} \bibnamefont{Gupta}},
  \bibinfo{journal}{Proc. Phys. Soc. A} \textbf{\bibinfo{volume}{65}}
  (\bibinfo{year}{1952}).

\bibitem[{\citenamefont{Law}(1995)}]{lawCavity}
\bibinfo{author}{\bibfnamefont{C.~K.} \bibnamefont{Law}},
  \bibinfo{journal}{Phys. Rev. A} \textbf{\bibinfo{volume}{51}}
  (\bibinfo{year}{1995}).

\bibitem[{\citenamefont{Cohen-Tannoudji
  et~al.}(1989)\citenamefont{Cohen-Tannoudji, Dupont-Roc, and
  Grynberg}}]{tannoudjiPhotons}
\bibinfo{author}{\bibfnamefont{C.}~\bibnamefont{Cohen-Tannoudji}},
  \bibinfo{author}{\bibfnamefont{J.}~\bibnamefont{Dupont-Roc}},
  \bibnamefont{and} \bibinfo{author}{\bibfnamefont{G.}~\bibnamefont{Grynberg}},
  \emph{\bibinfo{title}{Photons and Atoms: Introduction to Quantum
  Electrodynamics}} (\bibinfo{publisher}{John Wiley and Sons},
  \bibinfo{year}{1989}).

\bibitem[{\citenamefont{Walls and Milburn}(2008)}]{wallsMilburn}
\bibinfo{author}{\bibfnamefont{D.~F.} \bibnamefont{Walls}} \bibnamefont{and}
  \bibinfo{author}{\bibfnamefont{G.~J.} \bibnamefont{Milburn}},
  \emph{\bibinfo{title}{Quantum Optics}} (\bibinfo{publisher}{Springer},
  \bibinfo{year}{2008}).

\end{thebibliography}
\end{document}